\documentclass[pdflatex,sn-mathphys-num]{sn-jnl}% Math and Physical Sciences Numbered Reference Style 
\usepackage{graphicx}%
\usepackage{multirow}%
\usepackage{amsmath,amssymb,amsfonts}%
\usepackage{amsthm}%
\usepackage{mathrsfs}%
\usepackage[title]{appendix}%
\usepackage{xcolor}%
\usepackage{textcomp}%
\usepackage{manyfoot}%
\usepackage{booktabs}%
\usepackage{algorithm}%
\usepackage{algorithmicx}%
\usepackage{algpseudocode}%
\usepackage{listings}%
\usepackage{comment}
\usepackage{dcolumn} %Align table columns on decimal point
\usepackage{bm}% bold math
\usepackage{float}
\usepackage{caption}

\theoremstyle{thmstyleone}%
\theoremstyle{thmstyletwo}%

\theoremstyle{thmstylethree}%

\begin{document}

%\title[Confinement-induced proliferation of vortices around ciliated marine larvae with complex morphologies]{Confinement-induced proliferation of vortices around ciliated marine larvae with complex morphologies}

%OLD TITLE \title[Confinement-induced proliferation of vortices around marine invertebrate larvae]{Confinement-induced proliferation of vortices around marine invertebrate larvae } 

\title[A universal hydrodynamic transition in confined marine invertebrate larvae]{A universal hydrodynamic transition in confined marine invertebrate larvae}

\author[1]{\fnm{Bikram} \sur{D. Shrestha}}
\equalcont{These authors contributed equally to this work.}

\author[1]{\fnm{Santhan} \sur{Chandragiri}}
\equalcont{These authors contributed equally to this work.}

\author[1]{\fnm{Christian} \sur{D. Gibson}}

\author[1]{\fnm{Nina} \sur{R. Couture}}

\author[1]{\fnm{Melissa} \sur{Ruszczyk}}

\author*[1,2,3]{\fnm{Vivek} \sur{N. Prakash}}\email{vprakash@miami.edu}

\affil[1]{\orgdiv{Department of Physics}, \orgname{College of Arts and Sciences, University of Miami}, \orgaddress{\city{Coral Gables}, \state{FL}, \country{USA}}}

\affil[2]{\orgdiv{Department of Biology}, \orgname{College of Arts and Sciences, University of Miami}, \orgaddress{\city{Coral Gables}, \state{FL}, \country{USA}}}

\affil[3]{\orgdiv{Department of Marine Biology and Ecology}, \orgname{Rosenstiel School of Marine, Atmospheric, and Earth Science, University of Miami}, \orgaddress{ \city{Miami}, \state{FL}, \country{USA}}}

\abstract{The ocean is teeming with a myriad of submillimeter-sized invertebrate planktonic larvae~\cite{sardet2020plankton}, which thrive in a viscous fluid environment~\cite{lauga2009}. Many of them rely on ciliary beating to generate fluid flows for locomotion and feeding~\cite{lauga2009,gilpin2020}. Their larval forms, local morphologies, and ciliation patterns exhibit remarkable diversity~\cite{emlet1991diversity}, producing intricate and dynamic three-dimensional (3D) flows that are notoriously difficult to characterize in laboratory settings. Traditional microscopic imaging techniques typically involve gently squeeze-confining the soft larvae between a glass slide and cover slip to study their flows in quasi-two-dimensions (2D)~\cite{gilpin2017NatPhys}. However, a comprehensive hydrodynamic framework for the low-to-intermediate Reynolds number ($<\sim$1) flows at the larval scale in quasi-2D confinement — particularly in light of their complex forms — has remained elusive. Here, we demonstrate that vortices around larvae proliferate with increasing confinement and illuminate the underlying physical mechanism. We experimentally quantify confinement-induced flows in larvae of sea stars and sea urchins. The flows exhibited strikingly universal patterns: under weak confinement, all larvae generated two vortices, whereas under strong confinement, the number of generated vortices significantly increased. The experimental observations were well captured by a low Reynolds number theoretical model based on the superposition of confined Stokeslets~\cite{liron1976,mondal2021}. Building on experiments and theory, we developed a comprehensive framework for confinement-induced larval flows, which suggests that vorticity dynamics are primarily determined by local morphological features, rather than solely the body plan. Our work provides fundamental insights into the form-functional relationships between larval morphology and flow generation. Our findings are broadly applicable to understanding fluid flows generated by a wide range of ciliated organisms with complex forms and morphologies, from micro- to milli-length-scales.}

\keywords{Biomechanics, Fluid dynamics, Marine invertebrate larvae, Confinement, Ciliary flows}

\maketitle

Diverse and exotic life forms inhabit the vast oceans and hold many secrets to life that are waiting to be unraveled~\cite{sardet2020plankton}. The biomechanics of marine organisms and their interactions with the fluid environment are fundamental drivers of ecological and evolutionary processes in the ocean. Marine species generate fluid flows for a variety of purposes, including  locomotion, feeding, predator evasion, reproduction, and development~\cite{denny1993,lauga2009,vogel2020,dabiri2019,wan2023}. Many benthic marine invertebrates, such as echinoderms (e.g. sea stars and sea urchins) undergo an indirect mode of development that involves a pelagic larval stage with a body plan completely different from their adult form. During this larval stage, these organisms swim and feed in the water column for weeks or even months before metamorphosing into their adult shape. The majority of these larvae are soft-bodied, planktonic, free-swimmers that utilize ciliary beating to create fluid flows for both swimming and feeding~\cite{gilpin2020,gilpin2017NatPhys,byron2021}. In these larvae, large numbers of motile cilia are arranged in ciliary bands embedded in their body morphologies enabling them to generate unexpectedly rich fluid-flow patterns. So far, sea star larvae have been shown to generate a fascinating array of multiple counter-rotating vortices ~\cite{gilpin2017NatPhys,gilpin2017dynamic}. Further, the diversity of ciliated larval forms is staggering, with remarkable variations in body morphology and ciliation patterns~\cite{emlet1991diversity}. Therefore, an open question is whether the differences in organism body plans and/or local morphologies between ciliated larvae give rise to distinct flow patterns. We will investigate this question for the first time by examining morphologically distinct larvae from two closely related echinoderms: the sea star (\textit{Patiria miniata}) and sea urchin (\textit{Lytechinus variegatus}). Specifically, we will investigate two developmental stages (early stage and late stage) of sea star larvae, and one stage of sea urchin larvae. Sea star larvae have a smooth oval/ellipsoid body plan, with bilateral and anterior-posterior symmetry; the early stage bipinnaria larvae have a simpler body plan compared to the more complex-shaped late stage brachioleria larvae, which feature local morphological variations. In stark contrast, the echinopluteus sea urchin larvae have a cone-shaped body plan, with rigid, extended skeletal arms and exhibit bilateral symmetry but lack anterior-posterior symmetry. In this study, we aim to experimentally quantify and model the flow patterns generated by these three distinct larval types, to uncover the fundamental relationships between the different forms of marine invertebrate larvae and their hydrodynamic signatures.

The tiny, sub-millimeter larvae reside in a viscous fluid environment in the ocean, where viscous forces dominate over inertial forces. This balance is quantified by the Reynolds number ($Re$), which represents the ratio of inertial to viscous forces, and is defined as $Re = UL/\nu$, where $U$, $L$, and $\nu$ are the characteristic velocity, length scale, and kinematic viscosity of water, respectively. For sea star and sea urchin larvae swimming in seawater, $Re\sim$ 0.1--0.9 (Table S1, Supplementary Information), placing them within a low-to-intermediate Reynolds number regime, where although viscous effects are prominent, inertial effects begin to play a role. Low-to-intermediate Reynolds number flows represent a relatively under-explored regime in fluid dynamics, as the majority of the existing literature on biological fluid mechanics focuses on either low Reynolds number flows ($Re\sim$ $1\times10^{-5}$ to 0.1)~\cite{lauga2009,ishikawa2024} or intermediate to high Reynolds number flows ($Re>1$)~\cite{denny1993,vogel2020}. In this study, we present new experimental results at $Re\sim0.1-0.9$, and demonstrate that theoretical models based on low Reynolds number assumptions provide a qualitatively accurate description of the observed phenomena. 

The majority of aquatic microorganisms at low-to-intermediate Reynolds numbers swim  freely in three dimensions (3D) within open water bodies such as lakes and oceans. However, these microorganisms frequently encounter rigid boundaries or surfaces due to interactions with other living organisms or non-living structures, such as the air-water interface or other solid matter, including substrates at the bottom. Microorganism-boundary interactions have important biological implications in various natural contexts. For example, sessile, suspension-feeding ciliated organisms like \textit{Vorticella}~\cite{pepper2010,pepper2013}, \textit{Stentor}~\cite{shekhar2023}, and coral~\cite{shapiro2014,kiel2022} need to optimize nutrient availability in their immediate surroundings since they are unable to swim away to find nutrient sources. Similarly, many animals rely on internal ciliary-driven flows within organs such as human airways, brain ventricles and reproductive tracts to regulate vital physiological processes~\cite{kanso2024}. 

The marine invertebrate larvae examined in this study likely come across rigid boundaries during terminal encounters or in laboratory settings. Nevertheless, they serve as an excellent model system for investigating the hydrodynamic effects of microorganism-boundary interactions. Confinement-induced fluid dynamical phenomena are often found to be highly non-intuitive and non-trivial, garnering considerable interest in the field of biological fluid dynamics, particularly at low Reynolds numbers~\cite{liron1976,emlet1990tether,catton2007,pepper2010,pepper2013,mathijssen2016,gilpin2017NatPhys,vonDassow2017,gilpin2017reply,jeanneret2019,mondal2021,bentley2022,bondoc2023,shekhar2023,selvan2023,radhakrishnan2024}. In the present work, we investigate confinement effects in low-to-intermediate Reynolds number flows for the first time.

The complex 3D flow fields generated by ciliated marine invertebrate larvae pose many challenges to quantification in laboratory-based experimental studies, despite recent advancements~\cite{krishnamurthy2020}. The straightforward traditional microscopic imaging methods, which confine larvae between a glass slide and a cover slip, impose a quasi-two-dimensional (2D) confinement. So far, this conventional quasi-2D approach has been instrumental in revealing phenomena such as the multiple vortex arrays generated by sea star bipinnaria larvae~\cite{gilpin2017NatPhys}. However, alternative confinement techniques, such as tethering larvae~\cite{vonDassow2017}, fail to produce the same multiple vortex patterns~\cite{gilpin2017NatPhys, gilpin2017reply, gilpin2017dynamic}. This discrepancy raises fundamental questions about the impact of quasi-2D confinement on the flow fields of 3D micro-swimmers. Here, we resolve this conundrum and address several longstanding questions that have puzzled biological fluid mechanicians for decades: How does quasi-2D confinement alter the fluid flow fields of a 3D micro-swimmer? Do variations in larval body plan and local morphology of ciliated larval micro-swimmers give rise to different flow patterns under quasi-2D confinement? Is there a unified theoretical hydrodynamic framework that can capture these quasi-2D confinement effects at low-to-intermediate Reynolds numbers? 

We address these open questions in physics and biology by investigating the effects of squeeze-confinement on the low-to-intermediate Reynolds number flows generated by three distinct forms of ciliated marine invertebrate larvae of echinoderms: (i) early stage sea star, (ii) late stage sea star, and (iii) sea urchin (Fig.~\ref{fig:1}). We systematically vary the height between the glass slide and cover slip ($H$) and quantify changes in the vorticity dynamics. We quantitatively compare our experimental results with a theoretical low Reynolds number-based model, and develop a universal hydrodynamic framework for confinement flows.  

\subsection*{Confinement-induced flows in marine larvae} 

\begin{figure}[H]
	\centering
  \includegraphics[width=1.2\columnwidth]{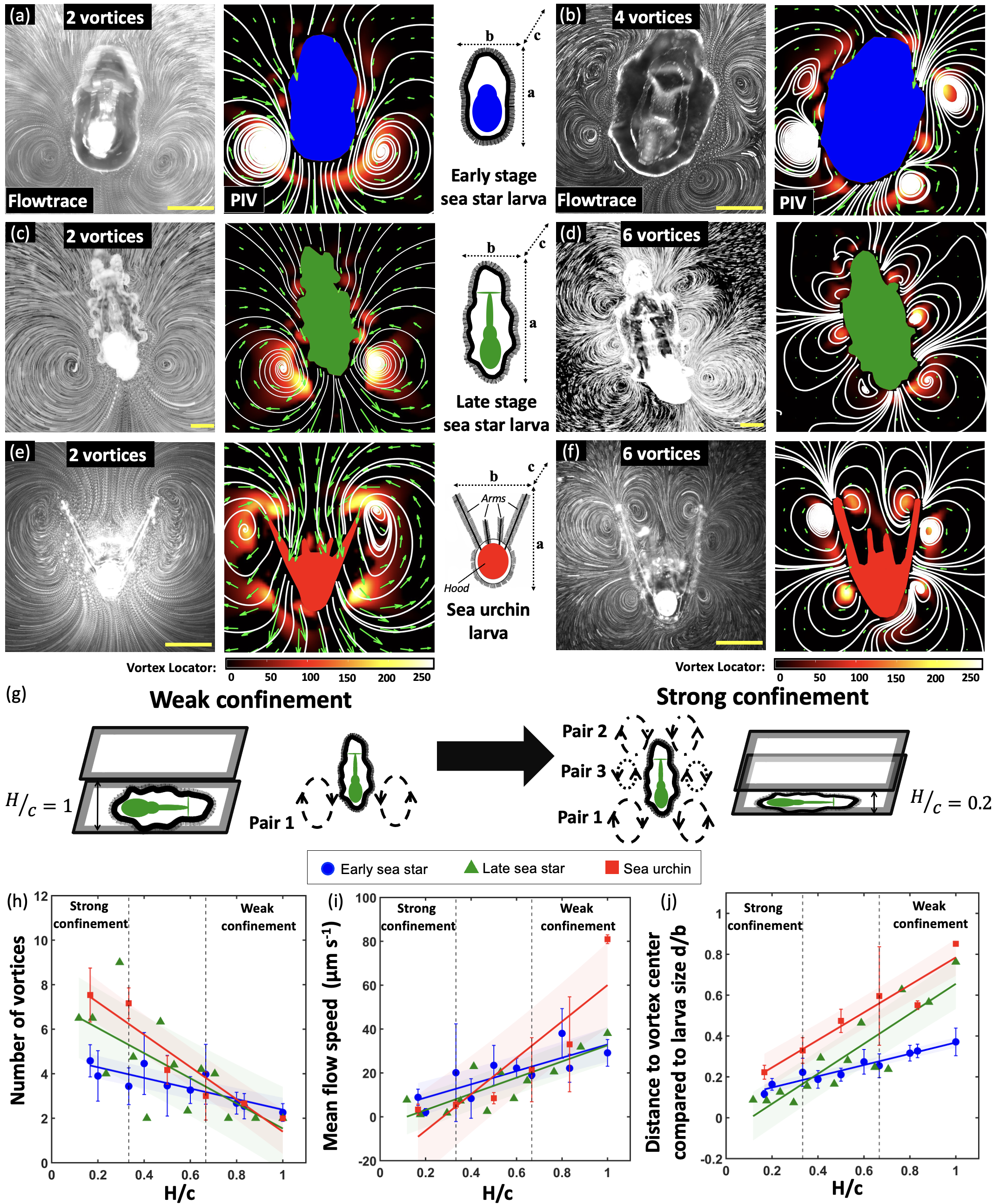} 
            \caption{(Caption next page.)}
		\label{fig:1}
\end{figure}

Using live darkfield time-lapse microscopy experiments, we observed that flow fields generated by the three distinct larvae varied depending on the degree of squeeze-confinement (Methods, Fig.~\ref{fig:1}). The larvae were confined by gently trapping or squeezing them between a glass slide and a cover slip that were separated by a precisely known height, $H$. This height was varied over a broad range for the three larvae -- from 'weak confinement' (low squeeze/large $H$) to 'strong confinement' (high squeeze/small $H$) and their resultant flow fields were quantified using standard techniques from experimental fluid dynamics (Fig.~\ref{fig:1}, Methods). The flow visualizations in Fig.~\ref{fig:1} (a-f) were carried out by post-processing of the time-lapse images using Flowtrace in ImageJ ~\cite{gilpin2017Flowtrace} (left panels), and Particle Image Velocimetry (PIV) in MATLAB~\cite{stamhuis2014pivlab} (right panels) (Methods). We observed that weak confinement (low squeeze/large $H$) resulted in two vortices in all the three types of larvae (Fig.~\ref{fig:1}a, c, e) (Video S1). When the confinement was increased by bringing the glass slide and cover slip closer, first reaching 'moderate' confinement conditions (medium squeeze/medium $H$) (Extended Data Fig.~\ref{fig:S1}) and then going to 'strong' confinement conditions (high squeeze/small $H$), all the larvae generated more than two vortices (Fig.~\ref{fig:1}b, d, f) (Video S2). In our quasi-2D experiments, the thin imaging z-plane was focused close to the mid-plane ($\sim H/2$) located in between the lower (glass slide) and upper (cover slip) boundaries. We found that there was minimal flow field variation in the z-planes above and below the mid-plane (Extended Data Fig.~\ref{fig:S2}). Hence our experiments captured the dominant and representative quasi-2D flow fields generated by the larvae.

\addtocounter{figure}{-1}
\begin{figure} [t!]
  \caption{(Previous page.) {\textbf{Fluid flow vortices generated by ciliated marine larvae in quasi-2D squeeze confinement experiments.}} (a, b): Early stage sea star larvae generate two vortices under weak confinement (a), and four vortices under strong confinement (b). (c, d): Late stage sea star larvae generate two vortices under weak confinement (c), and six vortices under strong confinement (d). (e, f): Sea urchin larvae generate two vortices under weak confinement (e), and six vortices under strong confinement (f). In a-f, the left panel shows flowtrace visualizations~\cite{gilpin2017Flowtrace} and the right panels display corresponding flow fields obtained using Particle Image Velocimetry (PIV)~\cite{stamhuis2014pivlab}. PIV panels show streamlines (white) superimposed with velocity vector field (green arrows), larval masks (blue, red, or green, depending on body plan), and vortex locator heat map. Throughout a-f, organisms are measured such that $a$ is major axis, $b$ is minor axis, and $c$ is its vertical height. (g) Schematic representation of changes in the height between the glass slide and cover slip, $H$. Left panels show weak confinement (low squeeze) and the right panels show strong confinement (high squeeze), along with pairs of vortices generated. The maximum vertical height, $H$, tested corresponds to the vertical height of the organism, $c$, indicating no squeeze confinement. (h-j): Flow field quantification in different confinement strengths (weak, moderate and strong), displaying total number of vortices, mean $\pm$ SD flow speed, and distance from vortex center to larval body, as a function of squeeze confinement parameter ($H/c$), where $c$ is the larval size in the depth (z) dimension. The data points represent mean values and errorbars represent standard deviations. The length scale bars in a-f (yellow color) correspond to 0.2 mm. Statistical descriptions of linear regressions can be found in Supplementary Table S8.}
\end{figure}

To account for the size variation among the three types of larvae, we divide the confinement height ($H$) length scale with the larval depth dimension ($c$), to arrive at a normalized squeeze-confinement parameter $H/c$ (Supplementary Information). This non-dimensional confinement parameter provides a standardized quantification of the degree of squeeze-confinement, and hence enables comparison of flows between the three larval types. Confinement strength can then be described categorically, using the same $H/c$ ranges for each larvae to describe similarities and differences between larvae body plans under strong ($0 \leq H/c < \frac{1}{3}$), moderate ($\frac{1}{3} \leq H/c < \frac{2}{3}$), and weak ($\frac{2}{3} \leq H/c \leq 1$) confinement conditions. We found that as confinement weakened ($H/c$ increased), the number of generated vortices for all larvae, regardless of their body plan and local morphology, converged to 2 vortices (Fig.~\ref{fig:1}a, c, e, Supplementary Tables S8-S9). However, when confinement strength increased ($H/c$ decreased), the total number of vortices generated varied depending on the confinement strength and larva body plan (Fig.~\ref{fig:1}b, d, f, Supplementary Tables S9-12). Under strong confinement ($0 \leq H/c < \frac{1}{3}$) conditions, the simple-shaped, early stage sea star larva developed four total vortices, i.e., two symmetric pairs (Fig.~\ref{fig:1}b). Although the complex-shaped, late stage sea star and sea urchin larvae developed a total of six vortices under strong confinement ($0 \leq H/c < \frac{1}{3}$) (Fig.~\ref{fig:1}d, f), the vortex locations were not similar since these two larvae have completely different body plans and local morphologies (Supplementary Tables S9, S12). Meanwhile, under moderate confinement conditions ($\frac{1}{3} \leq H/c < \frac{2}{3}$), the three different types of larvae generated more than two vortices (Extended Data Fig.~\ref{fig:S1}), but less than the maximum number they generated under strong confinement (Fig.~\ref{fig:1}b, d, f, Supplementary Tables S9-S11). Taken together, these results suggest that both the confinement parameter and larval morphology play an important role in determining vortex dynamics under confinement.

Regardless of the larval body plan and local morphology, we observed that as $H/c$ increased the number of vortices converged to 2 for all larval body plans (Fig. \ref{fig:1}h). However, when the degree of squeeze-confinement increased ($H/c$ decreases), the number of vortices increased  (Fig.~\ref{fig:1}h, Supplementary Tables S8-11). Under the strongest observed squeeze-confinement conditions ($H/c\sim0.2$) (Fig.~\ref{fig:1}h), both the morphologically complex sea urchin larvae with extended arms, and the morphologically complex late stage sea star larvae developed a maximum of nine vortices, more than the simpler early stage sea star larvae, which only generated a maximum of six vortices (Fig. S1, Supplementary Information). In fact, there is a statistically significant difference between the number of vortices by sea urchin and early stage sea star larvae under strong confinement ($0 \leq H/c < \frac{1}{3}$, Supplementary Table S12). 

Overall, our results reveal that all three types of larvae generate two vortices under the weakest confinement ($H/c = 1$), and the number of vortices increase with confinement ($H/c$ decreases) following an almost linear trend despite variations in their body plan and local morphology (Fig.~\ref{fig:1}h, Supplementary Table S8). From a physical perspective, these results suggest that a greater number of vortices are generated due to increased frictional effects when the upper (cover slip) and lower (glass slide) boundaries come closer to each other. Hence, we believe that this trend of increasing vortex number with increased confinement is universal and is applicable to a wide majority of marine ciliated larvae at comparable sizes. The number of vortices generated are also affected by the larva's local morphology, with the most complex larval morphologies (sea urchin as the most complex, and early stage sea star as the least complex) generating significantly more vortices than simpler morphologies under strong confinement (Supplementary Tables S9-S10).

In addition to the total number of vortices generated, the larval flow fields under different confinement conditions showed systematic variations in other flow metrics such as the mean flow speeds generated (Fig.~\ref{fig:1}i) and the distance of the vortices from the organism  (Fig.~\ref{fig:1}j) (Methods, Supplementary Tables S9-12). We observed that the mean flow speeds were greatest under weak confinement ($\frac{2}{3} \leq H/c \leq 1$) and the speeds decreased almost linearly with increasing confinement ($H/c$ decreases) (Fig.~\ref{fig:1}i, Extended Data Fig.~\ref{fig:S3}c, Supplementary Tables S10-11). This trend can likely be attributed to an increase of frictional effects from the walls due to increased confinement, leading to suppression of the strength of fluid flows (speeds) generated by the larvae. 
 %It has been reported in the fluid dynamics literature that the strength of the flow decreases with an increase in the effect of the confinement (\textcolor{red}{cite accordingly}??). 
We also observed that sea urchin larvae generate fluid flows of higher mean speeds compared to the sea star larvae under weak confinement (Fig.~\ref{fig:1}i), however, these results are not significantly different than the sea star larvae, likely due to uneven sample size (Supplementary Table S12). Next, our quantification of the distances from vortex centers to the larval body surface also showed a maximum value under weak confinement ($\frac{2}{3} \leq H/c \leq 1$), and a similar decreasing linear trend with increased confinement ($H/c$ decreased) (Fig.~\ref{fig:1}j, Extended Data Fig.~\ref{fig:S3}d, Supplementary Tables S8-S10). This result implies that the vortices approach the larval body surface as confinement increased ($H/c$ decreased). Finally, we also quantified the mean vorticity and circulation generated by the three larvae, and once again found a similar trend of linear decrease in values with increased confinement (Extended Data Fig.~\ref{fig:S2}e, f, Supplementary Tables S8-S11). Hence, all our experimental flow quantification results support our physical interpretation of increasing frictional effects from the walls due to increased confinement. Overall, these experimental results suggest the existence of an underlying hydrodynamic mechanism that determines the vortical flow patterns under different confinement conditions.

\subsection*{Experimental and Theoretical Flow Quantification} 

To investigate the hydrodynamics of confinement-induced flow fields around marine ciliated larvae, we quantified the vorticity in the experimental data and developed a theoretical model that closely matches the experimental flow fields. The $Re$ for the three types of marine larvae studied here are in the range $Re \sim 0.1-0.9$ (Methods, Supplementary Table S1), which corresponds to a low-to-intermediate $Re$ range where the flows are more dominated by viscous forces compared to inertial forces. Hence, we adopted a recently developed low $Re$ hydrodynamics model~\cite{mondal2021} to quantitatively compare the experimental results. The model, based on the quasi-2D Stokes equation for low $Re$ flows consists of superposition of pairs of Stokeslets between two parallel flat plates, and explicitly takes into account the swimmer (larva) force $F$, and the chamber height $H$ (squeeze-confinement parameter) ~\cite{mondal2021,liron1976}. The final velocity field expressed in polar coordinates ($r,\phi$) is given by:
\begin{equation*}     %\label{eq:1}
    \begin{bmatrix}
        v_x \\
        v_y
    \end{bmatrix} 
    \left(r,\phi\right) = \frac{F}{2\pi^2\eta} \int_{-\pi/2+\phi}^{\pi/2+\phi} d\theta \int_{0}^{\infty} dk 
    \begin{bmatrix}
        \sin^2\theta \\
        \sin\theta \cos\theta
    \end{bmatrix} \frac{k \cos\left[kr\cos\left(\theta-\phi\right)\right]}{\left(k^2+\frac{\pi^2}{H^2}\right)}
\end{equation*}

where, $v_x,v_y$ are the velocity components in $x,y$ directions, and $\eta$ is the dynamic viscosity of the fluid. We input values for the force ($F$) and chamber height ($H$) corresponding to experiments into the theoretical model to compare the experimental flow-field results (Methods, Supplementary Information).

\begin{figure}[H]
	\centering
  \includegraphics[width=1.2\columnwidth]{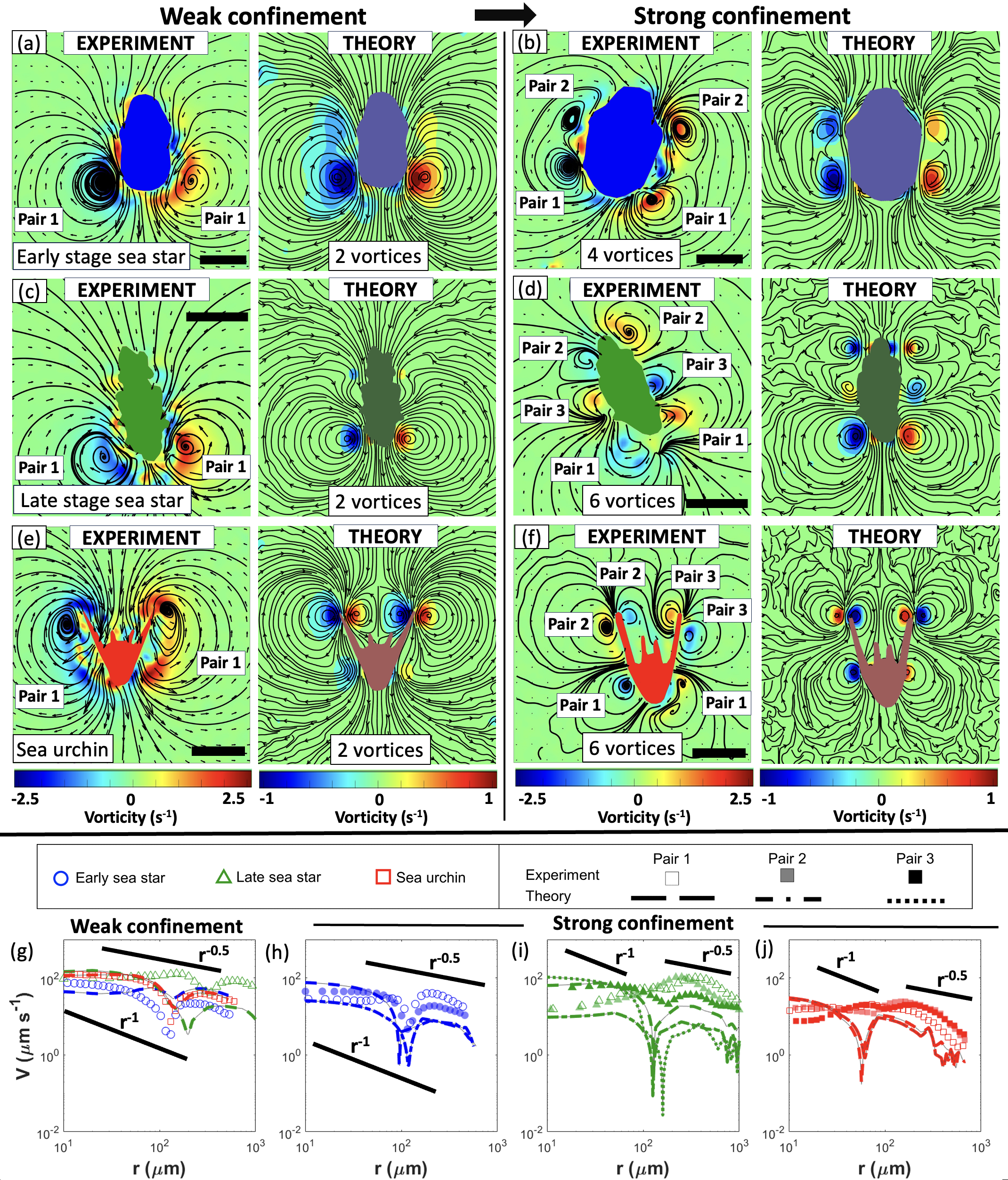} 
            \caption{(Caption next page.)}
		\label{fig:2}
\end{figure}

The comparison of the flow field results obtained from PIV-analysis of our experimental data~\cite{stamhuis2014pivlab,stamhuis2006} and our theoretical model are shown in Fig.~\ref{fig:2}. In weak confinement experiments ($\frac{2}{3} \leq H/c \leq 1$), all the larvae, despite their distinct body plans and local morphological variations, generated one dominant (large) pair of counter-rotating vortices (Pair 1) (Fig.~\ref{fig:2}a, c, e, left panels).

\addtocounter{figure}{-1}
\begin{figure} [t!]
  \caption{(Previous page.)  \textbf{Experimental and theoretical quantification of confinement-induced fluid flows in marine larvae.} (a-f) First, second and third rows show comparison of flow fields obtained from Experiments (PIV-derived vorticity) and Theory (superposition of Stokeslets) under weak and strong confinement for early stage sea star, late stage sea star, and sea urchin larvae (masked in blue/green/red colors, desaturated for theory). (a, c, e) Experiments and theory show only two vortices (Pair 1) being generated under weak confinement in all three types larvae. (b, d, f) Under strong confinement, early stage sea star larvae generate four vortices, and both late stage sea star and sea urchin larvae generate six vortices. The additional pairs of vortices (Pair 2 and 3) are also indicated. (g-j) Velocity quantification comparing experiments and theory: plots represent decay of velocity magnitude along a line perpendicular to the larval body surface and passing through the vortex center in Experiments (symbols) and Theory (lines). Velocity decay curves are represented by dashed (Pair 1), dashed-dotted (Pair 2), or dotted (Pair 3) lines. The solid black lines represent the velocity decay scaling v$\sim$$r^{-1}$ and v$\sim$$r^{-0.5}$. The length scale bars in Experimental panels (a-f) correspond to 0.2 mm.}
\end{figure}

In the oval-shaped, early stage sea star larva and more complex-shaped, late stage sea star larva under weak confinement, the pair of vortices are roughly circular-shaped, and their centers are located towards the back-end of the body (Fig.~\ref{fig:2}a, c, left panels). %Whereas, in the Sea urchin larva, we observed an extended pair of vortices, one on each side of the arms. 
The theoretical model utilized two pairs of Stokeslets for early stage sea star and three pairs of Stokeslets for the late stage sea star larvae under the same experimental confinement conditions ($H$) and physical location around the body surface (Methods, Supplementary Information), and showed excellent agreement with the experiments for both types of sea star larvae (Fig.~\ref{fig:2}a, c, right panels) (Extended Data Fig.~\ref{fig:S4}a, d). It must be noted that in the theoretical model, each Stokeslet generates 1 pair of vortices. However, the resultant velocity field due to the superposition of multiple Stokeslets gave rise to a flow field consisting of two vortices under weak confinement, closely matching the experiments (Extended Data Fig.~\ref{fig:S4}, Methods, Supplementary Information). 

In sea urchin larvae, weak confinement ($\frac{2}{3} \leq H/c \leq 1$)  resulted in vortices that are highly elongated in an ellipsoidal shape and distributed over the sides of the arms. Additionally, the vortex centers are located near the tip of the extended arms (Fig.~\ref{fig:2}e left panel). The theoretical model here utilized two pairs of Stokeslets under the same experimental confinement conditions ($H$) and physical location at the tip of the extended arms (Methods, Supplementary Information), and showed a flow field comparable in magnitude with the experiments, but not in the number of vortices generated (Fig.~\ref{fig:2}e, right panel). Compared to the sea stars, sea urchin morphology is much more complex, with 
with extended rigid arms. In this developmental stage, the sea urchin larva also had two small rudimentary arms near its oral surface, and these could also potentially influence the observed flow fields. For simplicity, the theoretical model does not account for the presence of these small arms. Hence, the theoretical flow fields are slightly different from that of experiments. Overall, each of the experimental larval vortices under weak confinement can be modeled theoretically with the superposition of multiple Stokeslets, and the experimental two-vortex larval flow field could be captured accurately. 

Under strong confinement conditions ($0 \leq H/c < \frac{1}{3}$), the local morphological variations between the three larvae play an important role in the number of vortices generated (Fig.~\ref{fig:2}b, d, f, left panels). While the early stage sea star generates two pairs of vortices (Pairs 1, 2) (Fig.~\ref{fig:2}b), both the late stage sea star and sea urchin larvae generate three pairs of vortices (Pairs 1, 2, 3) (Fig.~\ref{fig:2}d, f). Hence, strong confinement promotes the development of additional pairs of vortices in all the larvae (Fig.~\ref{fig:2}b, d, f). The early stage sea star larva has a smooth oval/ellipsoid-shaped body plan, and although it does not yet have arms protruding outside the body, there is a sharp convex curvature change in the region where arms will be formed in the future. The theoretical model with two pairs of Stokeslets placed at exactly the region corresponding to the convex curvature change shows excellent agreement with the experiments (Fig.~\ref{fig:2}b, right panel). 

In addition to the oval-shaped body, the late stage sea star larvae also have protrusions which act as localized sources of vorticity generation. Hence, under strong confinement, these larvae generate three pairs of vortices (Pairs 1, 2, 3) (Fig.~\ref{fig:2}d, left panel). The theoretical model considering three pairs of Stokeslets placed exactly adjacent to the these protrusions, once again shows excellent agreement with the experiments (Fig.~\ref{fig:2}d, right panel). 

In the sea urchin larvae, the tips of the extended arms and a hood region just below the arms also act as localized sources of vorticity generation. These local morphological complexities enabled larvae to generate three pairs of vortices (Pairs 1, 2, 3) under strong confinement (Fig.~\ref{fig:2}f left panel). To develop these three pairs of vortices in theory, a 2 pair Stokeslet model is used. In this context, one pair of Stokeslets are introduced at the back of the sea urchin larvae that accounted for the hood. Next, at the tip of each arm, one Stokeslet was introduced. Overall, this two pair Stokeslet model is able to capture the sea urchin larval experimental flow fields very well (Fig.~\ref{fig:2}f left panel).  

Overall, the theoretical model using superposition of Stokeslets qualitatively captured the experimental results very well in all three morphologically complex larvae under different confinement conditions, ranging from weak to high squeeze confinement. Further, the theoretical model was compared quantitatively with the experiments by calculating the magnitude of spatial velocity decay over the vortices (Fig.~\ref{fig:2}g-j). Under weak confinement conditions ($\frac{2}{3} \leq H/c \leq 1$) , since all the three larvae generate only one pair of vortices (Pair 1), we compared their velocity magnitude decay through the two vortex centers (averaged over Pair 1) from both the experiments and theoretical model in Fig.~\ref{fig:2}g. For the three types of larvae, the velocity at the larval body surface starts in the range 50-150 $\mu$m $s^{-1}$ and decays to a minimum (sharp dip representing the vortex center), and eventually decays to about half of its magnitude (30-70 $\mu$m $s^{-1}$) over a length scale of about $10^3$ $\mu$m (Fig.~\ref{fig:2}g). For the early stage sea star, the theoretical model matches the experiments almost exactly, and for the late stage sea star and sea urchin, the model is unable to match the experiments quantitatively, but still follows the overall trends. The velocity decays for all the types of larvae follow slopes in the range between $v \sim r^{-1}$ and $v \sim r^{-0.5}$. The initial velocity decay is closer to $v \sim r^{-1}$, and after crossing the velocity dip at the vortex center, the decay becomes less steep with a slope closer to $v \sim r^{-0.5}$ (Fig.~\ref{fig:2}g).      

For strong confinement ($0 \leq H/c < \frac{1}{3}$), the spatial velocity magnitude decays between the experiments and the theoretical model for the three types of larvae were compared for each vortex pair (Pair 1, 2, 3) separately (Fig.~\ref{fig:2}h-j). The theoretical model has the best match for the early stage sea star, followed by the sea urchin, and then the late stage sea star larvae. The experimental velocity decays from a range of 8-35 $\mu$m $s^{-1}$ at the surface to about half the value (3-15 $\mu$m $s^{-1}$) for the early stage sea star and sea urchin larvae (Fig.~\ref{fig:2}h, j), but this trend does not hold for all vortex pairs in the late stage sea star larvae (Fig.~\ref{fig:2}i). In all the three types of larvae, the theoretical model shows a steeper initial velocity decay of $v \sim r^{-1}$ compared to the experimental data, but the decay slope of $v \sim r^{-0.5}$ towards the end matches the experimental data (Fig.~\ref{fig:2}h-j). In all larvae, the model predicts a sharp dip in velocity at the vortex center, but only the early stage sea star experiments show this sharp dip (Fig.~\ref{fig:2}h). 

Our theoretical model captures the experimental velocity decay trends reasonably well, but not the magnitudes. For the three types of larvae, in both the weak and strong confinement conditions, although the theoretical model velocity decay slopes range between $v\sim r^{-1}$ and $v \sim r^{-0.5}$, the experimental data more closely follow the $v \sim r^{-0.5}$ decay (Fig.~\ref{fig:2}g-j). In comparison, the velocity decays in the low Reynolds numbers flows in \textit{Chlamydomonas reinhardtii} have been found to decay as $v\sim r^{-2}$~\cite{jeanneret2019}, which is also the expected velocity decay for a point source confined between two parallel no-slip walls~\cite{liron1976,jeanneret2019}. The less steep velocity decays observed in the present work, in the range between $v\sim r^{-1}$ and $v \sim r^{-0.5}$, can most likely be attributed to inertial effects since $Re\sim0.1-0.9$. Future theoretical work in this regime of low-to-intermediate Reynolds numbers ($Re\sim$0.1--0.9) is necessary to deepen our understanding of the physical mechanisms underlying the velocity decays.

\subsection*{Framework for confinement-induced vorticity proliferation}

The theoretical model described above using superposition of Stokeslets matched the experimental confinement-induced flows and revealed two key features of confinement-induced flows in the morphologically complex larvae of sea stars and sea urchins. First, under strong confinement ($0 \leq H/c < \frac{1}{3}$), the theoretical model utilized a unit Stokeslet to represent a localized source of vorticity generation in the experiments, corresponding to a morphological location on the larval body such as the convex curvature change in the early stage sea star larvae, the protrusions in late stage sea star larvae, or the tip of extended arms in sea urchin larvae (Fig.~\ref{fig:2}b, d, f). Second, a quantitative comparison of the experimental confinement-induced velocity decays for all the types of larvae (Fig.~\ref{fig:2}g-j) revealed that the velocity magnitudes under weak confinement ($\frac{2}{3} \leq H/c \leq 1$)  are about five times higher than the velocity magnitudes under strong confinement. This result suggests that frictional effects from the bottom and top walls decrease the velocity magnitude as the walls come closer due to the increase in confinement. Putting together these two key features, we present a general framework to explain the increase in number of vortices due to the increase in confinement.

\begin{figure}[H]
	\centering
		\includegraphics[width=0.98\columnwidth]{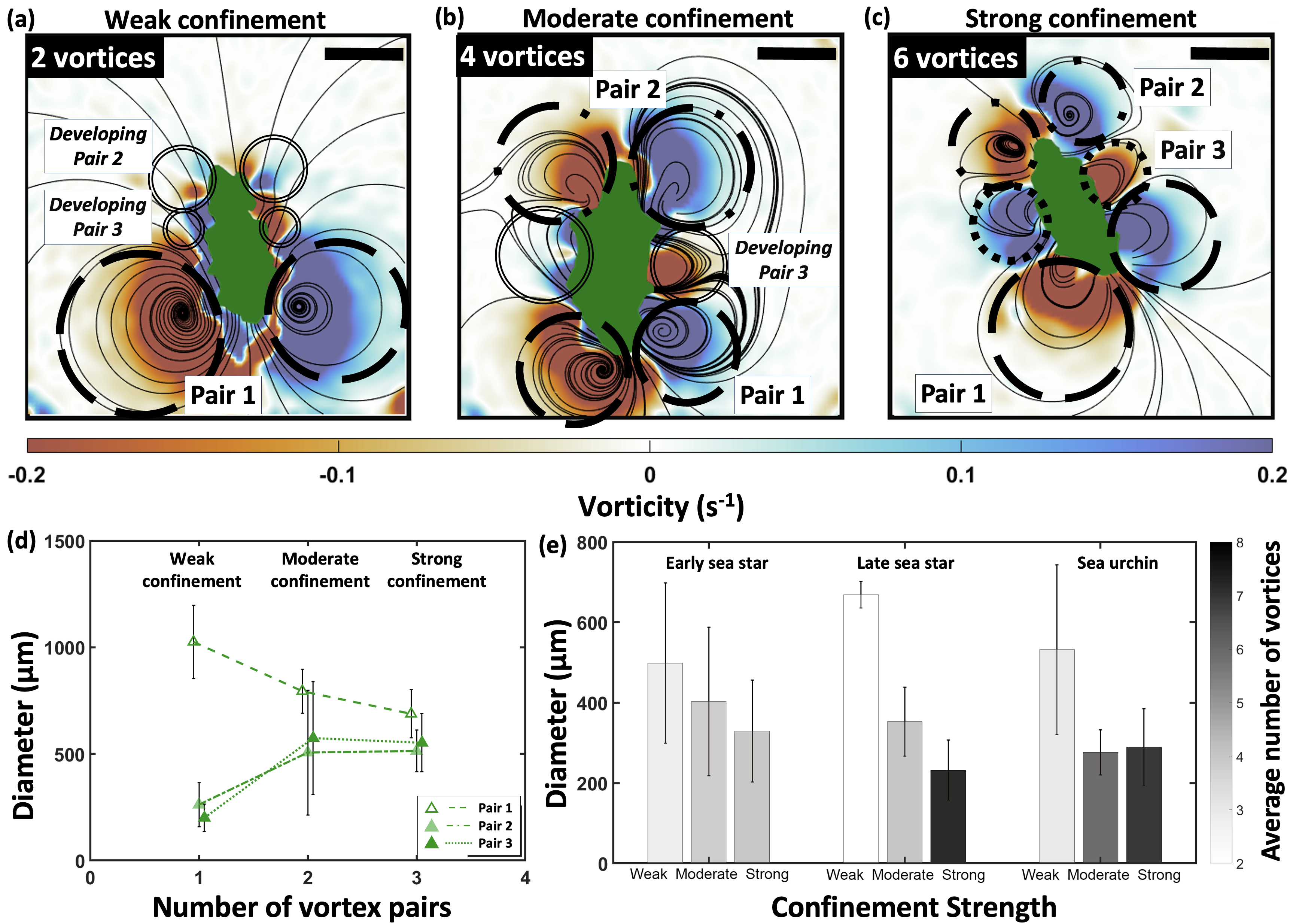}  
		\caption{\textbf{Local vorticity proliferation with increasing confinement in experiments.} (a-c) In late stage sea star larvae, the three pairs of vorticity sources around the larval body from experiments are shown under the different squeeze-confinement conditions (weak, moderate and strong). (a) Only vortex pair 1 is fully formed,(dashed circles). Weak regions of vorticity indicate where vortex pairs 2 and 3 will develop (double circles) (b) Vortex pairs 1 (dashed circles) and 2 (dashed-dotted circles) are fully developed; vortex pair 3 continues to develop. (c) All three vortex pairs are fully developed. (d) The vortex diameter across a pair averaged over different vorticity thresholds (Supplementary Table S2), versus the total number of vortex pairs generated around the late stage sea star larva. Increase in the number of vortex pairs corresponds to the increase in confinement (Fig.~\ref{fig:1}h). The dominant Pair 1 vortex size decreases with increase in confinement, whereas Pair 2 and 3 vortices increase in size with increase in confinement. Data points have been jittered about the x-axis for clarity. (e) The average diameter of all vortices for weak, moderate, and strong confinement for early sea star, late sea star, and sea urchin larvae. Bars are colored corresponding to the number of fully formed vortices. While panels a-d highlight this trend in one specific case of late stage sea star larvae, all organisms tested undergo a decrease in vortex size with respect to increasing confinement that results in a greater number of observed vortices. The length scale bars in panels (a-c) correspond to 0.5 mm.} 
		\label{fig:3}
\end{figure}

In the example of the late stage stage sea star larva, the body morphology consists of an extended oval/ellipsoidal shape with protruding arms. This late stage sea star larva generates two vortices under weak confinement ($\frac{2}{3} \leq H/c \leq 1$, Fig.~\ref{fig:1}c, Fig.~\ref{fig:2}c), four vortices under moderate confinement ($\frac{1}{3} \leq H/c < \frac{2}{3}$, Extended Fig.~\ref{fig:S1}), and six vortices under strong confinement ($\frac{2}{3} \leq H/c \leq 1$, Fig.~\ref{fig:1}d, Fig.~\ref{fig:2}d). In Fig.~\ref{fig:3}a-c, the vorticity field generated by the late stage sea star larva is reexamined for weak, moderate and strong confinements. The three source pairs (Pair 1, 2, 3) of local vorticity generation are visible in all the three confinement conditions. In weak confinement, the Pair 1 vortices dominate the flow-field with the largest vortex size (area) (Fig.~\ref{fig:3}d) (Extended Data Fig.~\ref{fig:S5}d) (Methods). However, the local sources of vorticity generation corresponding to the locations of the protruding arms (Pairs 2 and 3)  are visible as small patches of vorticity (Fig.~\ref{fig:3}a, d). In moderate confinement, the frictional effects from the top and bottom walls are increased, which decreases the size of the dominant Pair 1 vortices (Fig.~\ref{fig:3}b, d), and in turn provides an opportunity for Pair 2 and 3 vortices to increase in size (Fig.~\ref{fig:3}d). However, only Pair 2 vortices develop properly given that these are located far away from Pair 1. Hence, Pairs 1 and 2 dominate the flow field and suppress the formation of Pair 3 vortices (Fig.~\ref{fig:3}b). Under strong confinement, the wall frictional effects are further increased, and this further decreases the size of Pair 1 vortices (Fig.~\ref{fig:3}d), and enables the generation of Pair 3 vortices. Hence, the flow field is dominated by all the three vortex Pairs (1, 2, 3), giving rise to a total of six vortices. This relationship between vortex size, number, and degree of confinement -- two, large vortices under low confinement and many, small vortices under high confinement -- is present in all larvae tested, suggesting similar trends across different forms and local morphologies. (Fig.~\ref{fig:3}e) (Extended Data Fig.~\ref{fig:S4})(Extended Data Fig.~\ref{fig:S5}). It follows from this that the number of observable vortices depends on the number of sources of local vorticity generation that are either suppressed or amplified depending on if the confinement is weak or strong, independent of the overall larval body plan and only dependent on the local sources of vorticity generation (local morphological features such as tips of arms, sharp convex regions, etc.).

In summary, we investigated the flow fields induced by squeeze-confinement at low-to-intermediate Reynolds numbers ($<\sim$ 1) in the distinct larval forms of early stage and late stage sea star and sea urchin larvae. In all three cases, larvae generated two vortices under weak confinement, but under strong confinement they developed multiple vortices. A low Reynolds number-based theoretical model utilizing superposition of multiple Stokeslets showed good qualitative agreement with the experiments. Based on the experiments and theory, we constructed a framework to explain the universality of vortex dynamics under confinement, highlighting the importance of local morphological features inducing local vorticity generation.

%Hence, these fluid dynamics results will be applicable to a broader range of marine ciliated larvae with more complicated and intricate morphologies, and our theoretical framework could extend to microorganisms over a wide size-range, from micro- to meso-scales.   
%Our work broadly unifies the form-functional relationships between ciliated larval shape and flow generation. 
 %Discuss Shape vs physics generally, and new designs and AI generated robots etc

 Hydrodynamics resulting from microorganism-surface interactions are intriguing and often give rise to non-intuitive physical phenomena. While the present work elucidates the fluid dynamics associated with larval squeeze-confinement, alternative confinement methods—such as tethering~\cite{emlet1990tether,vonDassow2017,gilpin2017reply} or strong adhesive attachment~\cite{mazia1975}—could drastically alter the boundary conditions and lead to novel fluid dynamic effects. Also, the larval flows in the current work are in the regime of low-to-intermediate Reynolds numbers (0.1-0.9), so even though viscous effects are more important, inertial effects also need to considered. Although our theoretical model based on low Reynolds numbers~\cite{mondal2021,liron1976} performs well qualitatively, it does not work quantitatively since the model does not account for inertial effects. Hence, there is a need for the development of a new class of theoretical microswimmer models in this low-to-intermediate Reynolds numbers regime.   
  
 Our study primarily relied on a quasi-2D flow approximation, which is well-suited for the larval scale. However, we anticipate that the complex morphologies of larvae will give rise to intricate localized flows in 3D, which could play a crucial role in swimming and feeding dynamics. Additionally, we have assumed that squeeze-confinement does not substantially affect local ciliary beating in our experiments, given that ciliary movement is likely regulated at the level of individual cells. These considerations highlight important avenues for future research and underscore the broader significance of these confinement-induced phenomena for the larval biology and biological fluid dynamics community.

% i'm skipping discussion on temporal assumptions.. for space saving...we can mention it (discuss)
%for adhesive refs ask Athula/Deema?

Confinement-induced fluid dynamical effects on microorganisms can also have important biological consequences~\cite{gilpin2017NatPhys,shekhar2023}. Although it is unlikely that echinoderm larvae encounter squeeze-confinement conditions in the ocean, they are likely to interact with surfaces and these hydrodynamic interactions can influence their swimming and feeding behaviors~\cite{gilpin2017NatPhys}, which will be the focus of our future studies. Thus, while our fluid dynamics results are generally relevant to microswimmer-surface interactions, they are particularly applicable to quantitative laboratory-based studies involving microscopic confinement of a wide range of ciliated and motile marine microorganisms with varied forms and  morphologies~\cite{bentley2022,shekhar2023}, including collective phenomena that emerge when large numbers of larvae interact in close proximity~\cite{fakhri2022,bardfalvy2024}. Beyond this, there are several potential applications in both biology and engineering. For instance, our findings could aid in the development of quantitative assays for studying larval morphological changes in response to chemicals~\cite{lion2024}, as well as ecological assays to assess microplastics accumulation in marine zooplankton~\cite{law2014microplastics,botterell2019microplastics}. In engineering applications, these insights from ciliary flows could contribute to the design of microfluidic devices like lab-on-a-chip systems\cite{cui2023}, artificial water purification devices~\cite{rubenstein1977filtration,urso2023}, and morphologically optimized micro-robots for targeted drug delivery~\cite{dillinger2021starfishrobot}, environmental monitoring~\cite{soto2021}, and beyond.

\section*{Methods} \label{sec:expt_and_sim}

\subsection*{Laboratory cultures, larval spawning, and maintenance}

Adult sea stars (\textit{Patiria miniata}) were collected and shipped to the University of Miami (Miami, FL, USA) from the California coast (Marinus Scientific, CA). Ten male and ten female sea stars were kept separately in two 20 gal glass aquarium tanks. Aquariums were filled with seawater (Imagitarium pacific ocean water, Petco) and maintained at a temperature of $16^\circ$C using an external chiller unit. The tank water was continuously circulated using a pump and filtered. Tanks underwent monthly $90\%$ water changes. The sea stars were hand-fed cut pieces of supermarket-purchased shrimp once a month, with each adult animal consuming approximately 1/3 of a medium-sized shrimp.

In-vitro fertilization of sea star gametes were carried out in large glass ovenware dishes (3 qt) in seawater using standard protocols~\cite{barone2024}. The gametes were collected by making a 2 cm lateral incision on the ventral side of the sea star arm parallel to the lateral canal. The eggs collected from female sea stars were kept in 100 $\mu$M 1-methyladenine (1-MA) to mature eggs for 30-60 min. Then, freshly collected sperm from male sea stars were added (1:1000 dilution) and gently mixed in the seawater. After successful fertilization, visually confirmed by the presence of a fertilization envelope, the glass dishes containing growing embryos were placed in incubators maintained at a temperature of $16^ \circ$C. Water changes ($50\%$) in the glass dishes were carried out weekly, removing the buildup of waste.

Sea star larvae were fed with red algae (\textit{Rhodomonas lens}) and green algae (\textit{Dunaliella tertiolecta}), obtained from the National Center for Marine Algae and Microbiota (NCMA, Bigelow, ME). Algae were cultured in 500 mL Erlenmeyer flasks covered with aluminum foil containing filtered seawater (FSW) and nutrients (Micro Algae Grow, Florida Aqua Farms) in 1:100 dilution of nutrient media in FSW, and kept in an incubator maintained at a temperature of $16^\circ$C, under a 10:14 hour dark:light cycle using LED lighting. Algae flasks were shaken at least once daily to prevent settling at the bottom. The algae from these flasks were fed to the larvae in 1:500 concentrations (algae:seawater). The algae cultures were split weekly by diluting the algae culture into 4 parts by volume of fresh seawater and nutrients.  

The sea urchin larvae (\textit{Lytechinus variegatus}) 4-5 days post-fertilization were obtained in 100 mL falcon tubes from the animal culture facility at The National Resource for Aplysia on the campus of University of Miami’s Rosenstiel School of Marine, Atmospheric, and Earth Science. The larvae were transferred to 1 L beakers with seawater, and were kept in the lab for 1-2 days during experimentation. They were fed red algae (\textit{R. lens}) and green algae (\textit{D. tertiolecta}), and the beakers were maintained at room temperature ($24^ \circ$C), with constant lighting.

\subsection*{Larval mounting and live imaging experiments}

For experiments, a randomly chosen larva was confined or squeezed gently between a bottom glass slide and a top cover slip with different heights (z-direction) using double-sided spacer tape of precise thickness (Nitto Inc., PET-based Double-coated Tape No. 5605, 5615, UTD-10B), and fluid flow-fields were quantified in a quasi-2D x-y plane. Slides were prepared by attaching a pair of long rectangular strips of double-sided tape of different thicknesses 50/100/150 $\mu$m on the slide edges to serve as spacers for a required boundary height ($H$) (Figure ~\ref{fig:1}). The tested boundary height ($H$) values ranged from 50-850 $\mu m$ (Supplementary Table S1). 

A  50 $\mu$L sample of seawater containing the larva was transferred using a P200 micropipette from the culture dish into the center of the prepared glass slide with the spacer tape walls at the edges. For flow visualization, 10 $\mu m$ of a 1:100 dilution of tracer polystyrene microspheres (Polysciences Inc, cat. no. 07310-15) in seawater was added to the droplet with the larva and gently mixed. 

A cover slip was slowly introduced onto the glass slide containing the larva and microspheres and attached to the spacer tape walls. Live-imaging experiments were carried out by mounting this slide on an upright microscope (Zeiss Axio Imager M2). The microsphere movements were imaged using darkfield time-lapse imaging using a 10X objective to capture a 1.5 mm $\times$ 1.5 mm field of view. The time-lapse images were acquired at 10 fps for a duration of 30 s using a high-speed camera (Hamamatsu ORCA-Fusion Gen-III sCMOS) attached to the Zeiss microscope. Three recordings of each organism were carried out. The best quality videos (i.e. no bead clumping, minimum larval displacement, no larval contraction, and larvae located in the middle of field-of-view away from the walls) were selected for analysis (Supplementary Table S1).

\subsection*{Experimental data post-processing and analysis}

The larval flow fields were qualitatively visualized using flowtrace technique in Image-J \cite{gilpin2017Flowtrace}. The analysis was run for 30 s durations over a time-averaging window of 3 s. Flowtrace visualizations were carried out on all the experimental datasets, and these videos were used to count the number of vortices (number of closed circular pathlines) for each $H$ (Table S1, Supplementary Information). The number of vortices was noted for the entire 30 s time sequence of a single dataset. In instances where there was high variation in the total number of vortices over 30s, the average number of vortices was calculated as a weighted average over the total number of frames. Else, the average number of vortices was assumed to be the mode. Weighted average for number of vortices was used primarily with the early star data due to larval contractions and displacements. The data were aggregated by organism for later statistical analysis (Table S1, Supplementary Information).

The Particle Image Velocimetry (PIV) technique was adopted for quantification of the experimental larval flows. The time-lapse images were post-processed using PIVlab 2.62 in MATLAB R2020b~\cite{stamhuis2014pivlab}. The time series were first imported into PIVlab’s GUI via the 'time resolved' Image sequencing style. Image pre-processing was done by enabling the Contrast Limited Adaptive Histogram Equalization (CLAHE) of 64 pixels with auto contrast stretch. A mask was created to exclude regions where a larva was present, ensuring that only the flow region was analyzed. A Fast Fourier Transform (FFT) based cross-correlation algorithm was used to compute the velocity fields. For the interrogation window settings, a multi-pass cross-correlation algorithm was employed with an initial window size of 100×100 pixels, and a second pass of 50×50 pixels, with 50$\%$ overlap between windows to improve spatial resolution.
After the analysis, the datasets were converted to physical units by applying a calibration factor based on a known reference scale in the image. Post-processing of the vector fields was carried out by validation the vectors using a standard deviation threshold of 8, and a local median filter threshold of 3. Missing or invalid vectors were interpolated using local median filtering. 
Additionally, a velocity threshold was applied to discard unrealistic velocities exceeding the expected range based on the experiments.
 The final images were generated by applying 'smooth data' filter to the highest level and high pass vector field strength to the lowest level. The same settings were applied for all the image sequences and mean was calculated for a 3 s time window. The entire analysis was repeated for multiple datasets to ensure consistency and reliability. Flow parameters such as vorticity or velocity magnitude were plotted on a 2-D, x-y plane with velocity vector fields. The average vorticity and velocity magnitude were calculated over a time window of 3 s to obtain smooth fluid flow visualizations and quantification. Finally, the distance from the center of vortices to the larval body surface was also measured (Figure S2, Supplementary Information). 

\subsection*{Statistical Analysis}

 In instances where multiple recordings occurred over one individual, dependent variables were averaged with respect to the individual to ensure independent measurements. This assumed that recordings taken from the same individual were comparable. Dependent variables analyzed included number of vortices, distance from organism to vortex center ($d/b$), average vortex circulation, average vortex size, average vorticity, and average vortex flow speed. 
 
 Data was compared with respect to two factors -- body plan (early stage sea star, late stage sea star or sea urchin), and confinement ($H/c$). To observe initial trends with respect to confinement ($H/c$), data were linearly regressed with respect to $H/c$.
 
 For more detailed statements on the relationship between flow fields, body plan, and confinement, confinement -- measured on a continuous scale -- was divided into three strengths: strong (0 $\leq$ H/c $< \frac{1}{3}$), moderate ($\frac{1}{3} \leq$ H/c $< \frac{2}{3}$), and weak ($\frac{2}{3} \leq$ H/c $\leq$ 1). Aggregating points allowed for organisms with different flow field patterns (i.e., number of vortices due to morphology such as early stage sea star and sea urchin) to be compared at similar confinement strengths. 

Data were not normally distributed and often violated the assumption of equal variance between factors (body plan and confinement strength) required for parametric analysis, and thus were analyzed using nonparametric means. Nonparametric Kruskal-Wallis tests and Steel-Dwass All-Pairs were used to analyze data with respect to body plan, confinement strength, confinement strength within body plan, and body plan within confinement strength. Statistical analysis was performed in JMP Pro 17.0 (SAS Institute Inc., Cary, NC). 

\subsection*{Theoretical model for confinement-induced flows} 

To theoretically model fluid flows generated by the ciliated marine larvae under squeeze-confinement, we used an approach utilizing superposition of Stokeslets ~\cite{mondal2021,liron1976}. In the experiments, ciliated larvae were confined between two plates and the flows were measured at the midplane, i.e. z = $H/2$. Recently, flow fields generated by a \textit{Chlamydomonas} confined between two parallel plates was studied using both experiments and theory~\cite{mondal2021}. Inspired by this work, we extended the theoretical model and applied it to our experiments in ciliated marine larvae. The theoretical model starts with the incompressible 3D Stokes equation (as in Ref~\cite{mondal2021}), 
\begin{equation}
    \nabla p\left(r\right)+\eta \nabla^2 v\left(r\right) = 0, \nabla.v\left(r\right) = 0.   
\end{equation}
where $p$ and $v$ are the fluid pressure and velocity fields respectively. Next, based on the quasi-2D Brinkman approximation, the equations are rewritten as 
\begin{equation}
    \nabla_{xy} p\left(r\right)+\eta \left(\nabla_{xy}^2- \frac{\pi^2}{H^2} \right) v\left(r\right)+F\delta\left(r\right) = 0, \nabla_{xy}.v\left(r\right) = 0.
\end{equation}
where $F$ is the point source of force generation in the z = $H/2$ plane. The solution to the above equation (using Fourier analysis) is obtained as:
\begin{equation}
    \begin{bmatrix}
        v_x \\
        v_y
    \end{bmatrix} 
    \left(r,\phi\right) = \frac{F}{2\pi^2\eta} \int_{-\pi/2+\phi}^{\pi/2+\phi} d\theta \int_{0}^{\infty} dk 
    \begin{bmatrix}
        \sin^2\theta \\
        \sin\theta \cos\theta
    \end{bmatrix} \frac{k \cos\left[kr\cos\left(\theta-\phi\right)\right]}{\left(k^2+\frac{\pi^2}{H^2}\right)}
\end{equation}
The velocity fields are numerically solved over a 30 x 30 x-y grid in MATLAB to obtain the theoretical flow fields. In Ref~\cite{mondal2021}, the \textit{Chlamydomonas} has two flagella that beat to generate vortices, however, here the marine larvae generate fluid vortices using a large number of cilia located on their ciliary bands. Although the methods of inducing fluid flow are different in the two studies, the consequence, i.e. vortex generation, is the same in both cases. Here, the focus is on modeling the overall flow fields induced by confinement rather than the local mechanism of vorticity generation. 

A unit Stokeslet is used to model the flow field corresponding to a single vortex. To account for the multiple vortices that are developed in the experiments, we superpose multiple Stokeslets (corresponding to each vortex) on the same physical locations on the larval body (determined from experiments) and obtain the resultant theoretical velocity field. Further details on the theoretical model, including the procedure for introduction of multiple Stokeslets for each type of larvae, and their corresponding flow fields are provided in the supplementary material.  

\bigskip

\backmatter

\bmhead{Supplementary information}

Supplementary information for this manuscript include the following items: 

PDF file with supporting text, Figures S1--S6, Tables S1--S12, and Video Legends. 

Videos S1 and S2.

\medskip

\bmhead{Acknowledgements}

We thank present and past Prakash lab members for their helpful suggestions and support. We thank Zak Swartz (Marine Biological laboratory) and Athula Wikramanayake, Nat Clarke, Christopher Roden (University of Miami) for help with sea star culturing and other helpful suggestions. We thank Phillip Gilette (University of Miami Rosenstiel School) for providing the sea urchin larvae. We thank Nitto Inc. (Atlanta, GA) for generously providing the double-sided spacer tapes used in our experiments. We thank Maciej Lisicki (University of Warsaw) and Lyndon Koens (University of Hull) for their invaluable suggestions on the theoretical modeling. V.N.P. thanks Manu Prakash (Stanford University) and William Gilpin (University of Texas at Austin) for many
insightful discussions over the years.

C.D.G. acknowledges funding support from the Florida-Georgia Louis Stokes Alliance for Minority Participation (FGLSAMP) undergraduate research program funded by the National Science Foundation (NSF) (grant $\#$0217675). V.N.P. acknowledges Start-up funding support and a Provost's Research Award from the University of Miami.

\medskip

\bmhead{Author Contributions} \hfill \break

Conceptualization: B.D.S., S.C., V.N.P.

Methodology - experiments: B.D.S., C.D.G., N.R.C.

Methodology - theory: S.C.

Data Analysis: B.D.S., S.C., M.R.

Writing – original draft: B.D.S., S.C., V.N.P.

Writing – review, editing: B.D.S., S.C., M.R., V.N.P.

Supervision: V.N.P.

Project Administration: V.N.P.

Funding Acquisition: V.N.P.

\medskip

\bmhead{Competing Interests} 
The authors do not have any competing interests. 

\medskip

\bmhead{Materials $\&$ Correspondence} 
Author to whom correspondence and materials requests should be addressed: V.N.P. E-mail: vprakash@miami.edu 

\medskip

\bmhead{Data availability} Datasets associated with the manuscript will be deposited to Zenodo (link to be provided after completion of peer review).

\medskip

\bmhead{Code availability} The computer codes that support the results from this study will be deposited to Zenodo (link to be provided after completion of peer review).

\medskip

\bibliography{sn-bibliography}% common bib file

\begin{comment}

\section*{Declarations}

Some journals require declarations to be submitted in a standardised format. Please check the Instructions for Authors of the journal to which you are submitting to see if you need to complete this section. If yes, your manuscript must contain the following sections under the heading `Declarations':

\begin{itemize}
\item Funding
\item Conflict of interest/Competing interests (check journal-specific guidelines for which heading to use)
\item Ethics approval and consent to participate
\item Consent for publication
\item Data availability 
\item Materials availability
\item Code availability 
\item Author contribution
\end{itemize}

\end{comment}

\newpage

%%%%%%%%%%%%%%%%%%%%%%%%%%%%%%%%%%%% EXTENDED FIGURES %%%%%%%%%%%%%%%%%%%%%%%%%%%%%%%%%%%%%%%%%%%%%%%%%%%%%%
\setcounter{figure}{0}

\begin{figure}%[b]
	\centering
		\includegraphics[width=0.7\columnwidth]{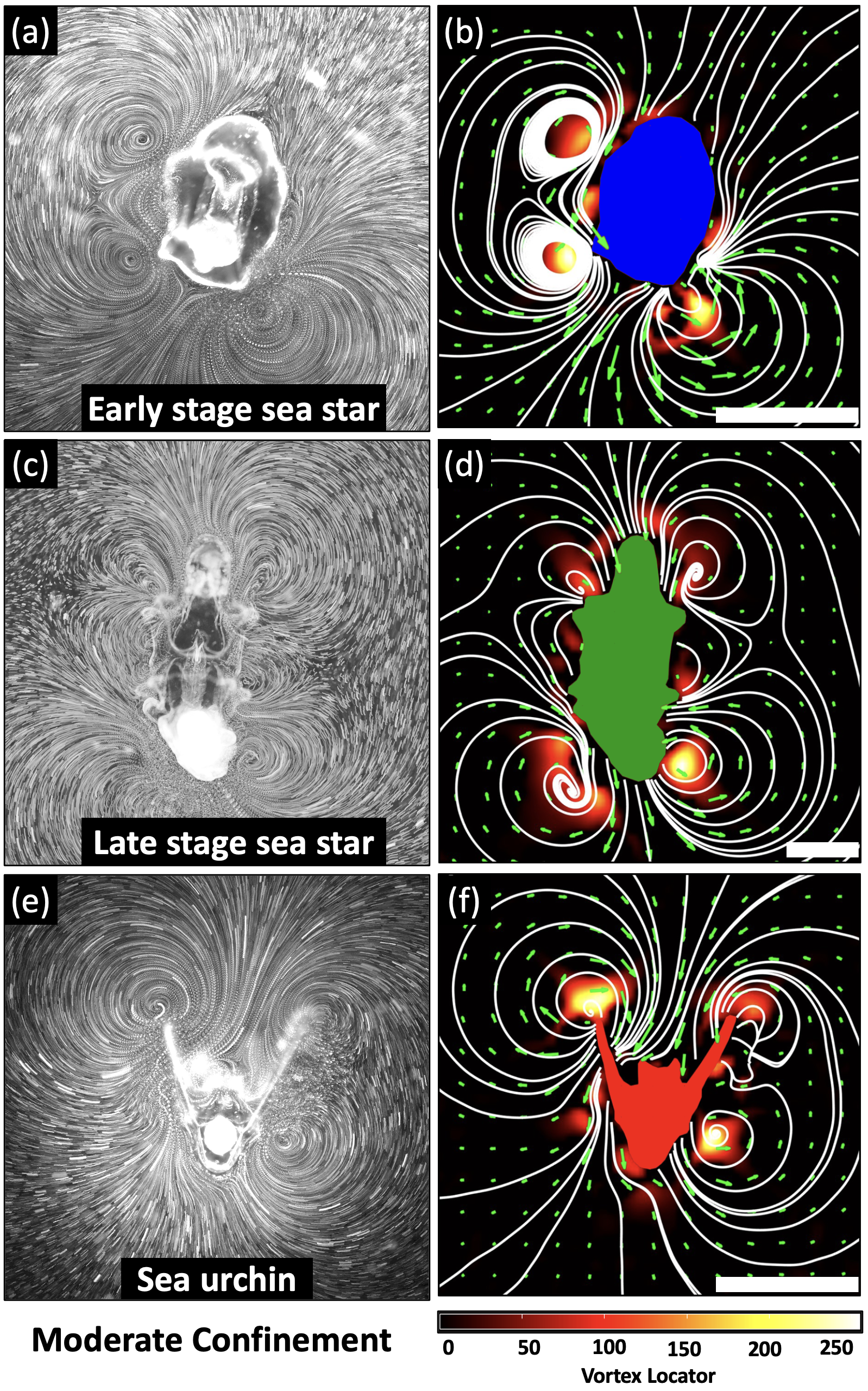} 	
        \captionsetup{labelformat=empty}
        \caption{\textbf{Extended Data Fig. 1: Fluid flow vortices around ciliated marine larvae under moderate squeeze confinement in experiments.} Left panels show Flowtrace visualization of fluid flow around the larvae, and right panels show the PIV-derived streamlines (white) along with the velocity vector field (green arrows) that are superimposed on the vortex locator (color contour). (a) An early stage sea star larva creates 3 vortices under moderate confinement. (b) A late stage sea star larva creates 4 vortices under moderate confinement. (c) A sea urchin larva creates 3 vortices under moderate confinement. The length scale bars (white) correspond to 0.4 mm.}
        \label{fig:S1}
\end{figure}

\begin{figure}
	\centering
		\includegraphics[width=0.9\columnwidth]{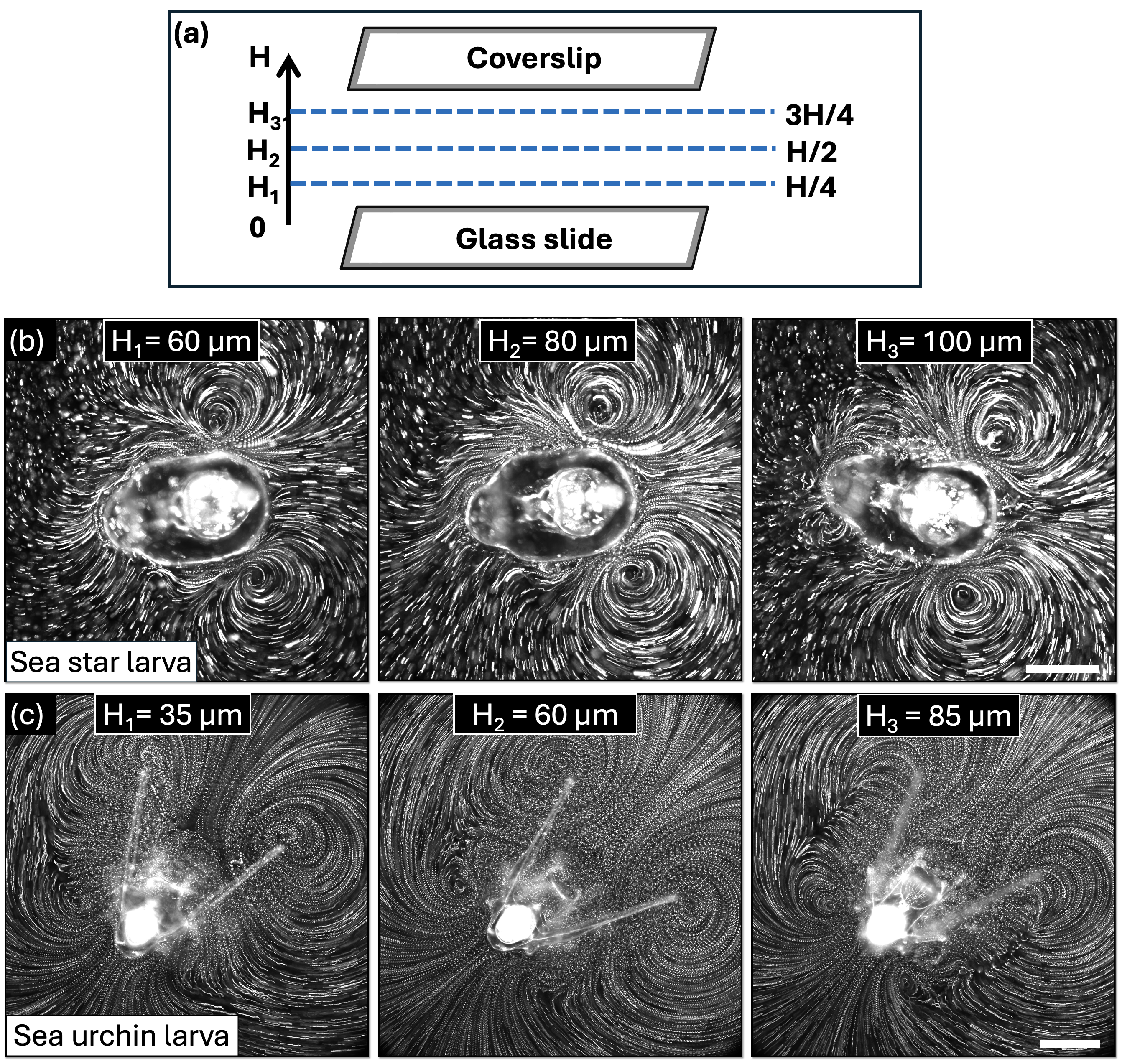} 
        \captionsetup{labelformat=empty}
        \caption{\textbf{Extended Data Fig. 2: Fluid flow-fields around squeeze-confined marine larvae at different z-planes.} (a) Schematic representation of the different z-planes of imaging between a glass slide and a cover slip on a microscope. (b) Flowtrace visualization around an early stage sea star larva at three z-planes (chamber height, $H$ = 150 $\mu$m). (c) Flowtrace visualization around a sea urchin larva at three z-planes (chamber height, $H$ = 100 $\mu$m). The length scale bars in (b, c) (white) correspond to 0.2 mm.}
		\label{fig:S2}
\end{figure}

\begin{figure}
	\centering
		\includegraphics[width=0.9\columnwidth]{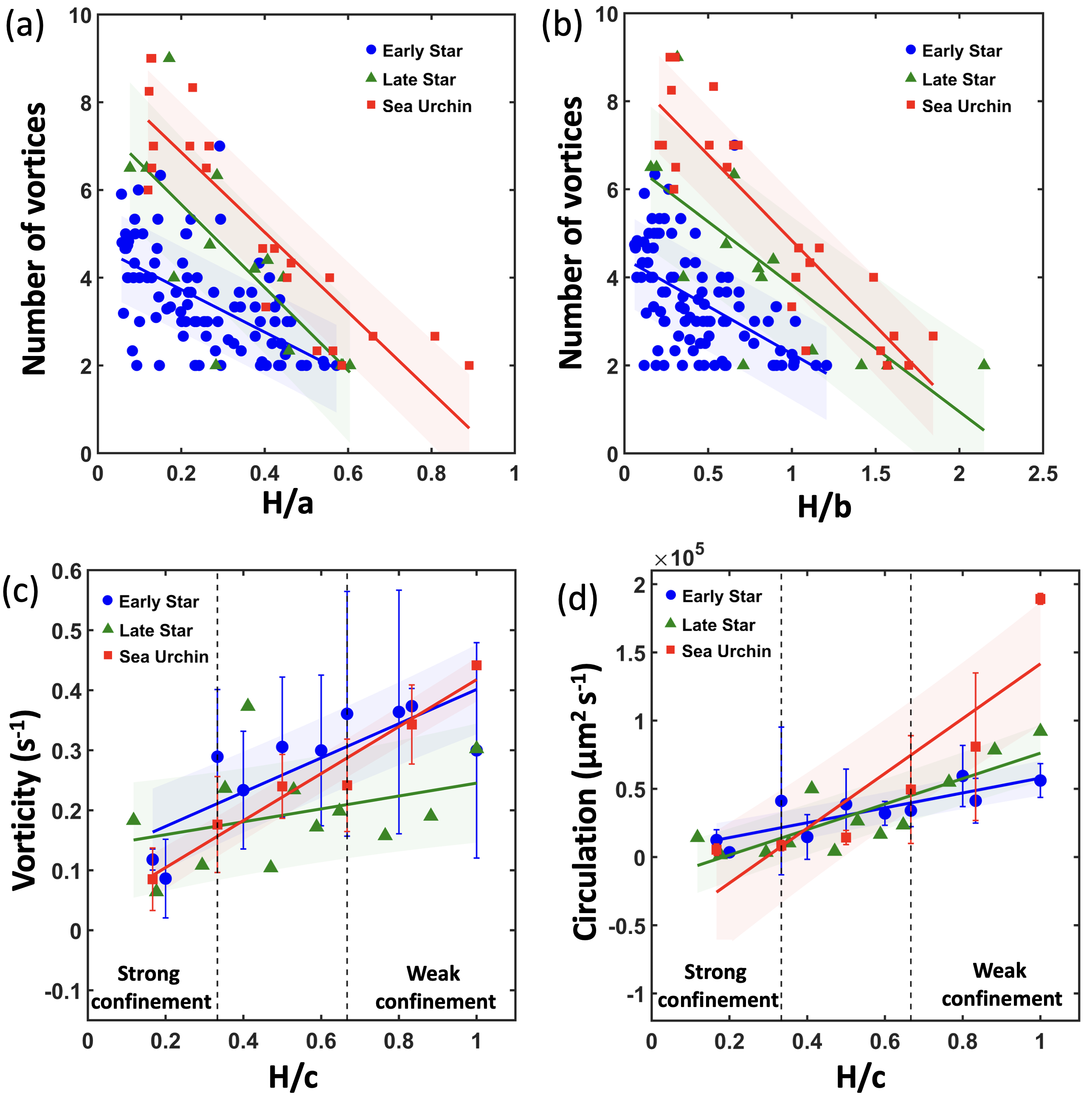}
        \captionsetup{labelformat=empty}
        \caption{\textbf{Extended Data Fig. 3: Quantification of flow parameters in experiments}. (a,b) Scatter plots of the total number of vortices generated by early stage sea star, late stage sea star, and sea urchin larvae as a function of normalized heights $H/b$ (a) and $H/a$ (b). Here, $b$ and $a$ are the length scales of larval height and width (schematic cartoons in Fig. \ref{fig:1}). The dotted lines are linear fits to the data for each type of larva (Supplementary Table S8). (c,d) The average vorticity and average circulation is plotted as a function of the normalized squeeze parameter ($H/c$) for the three types of larvae. The data points indicate mean values and errorbars indicate standard deviations, and dotted lines indicate linear fits to the mean values for each type of larva (Supplementary Table S8).}
		\label{fig:S3}
\end{figure}

\begin{figure}
	\centering
  \includegraphics[width=0.9\columnwidth]{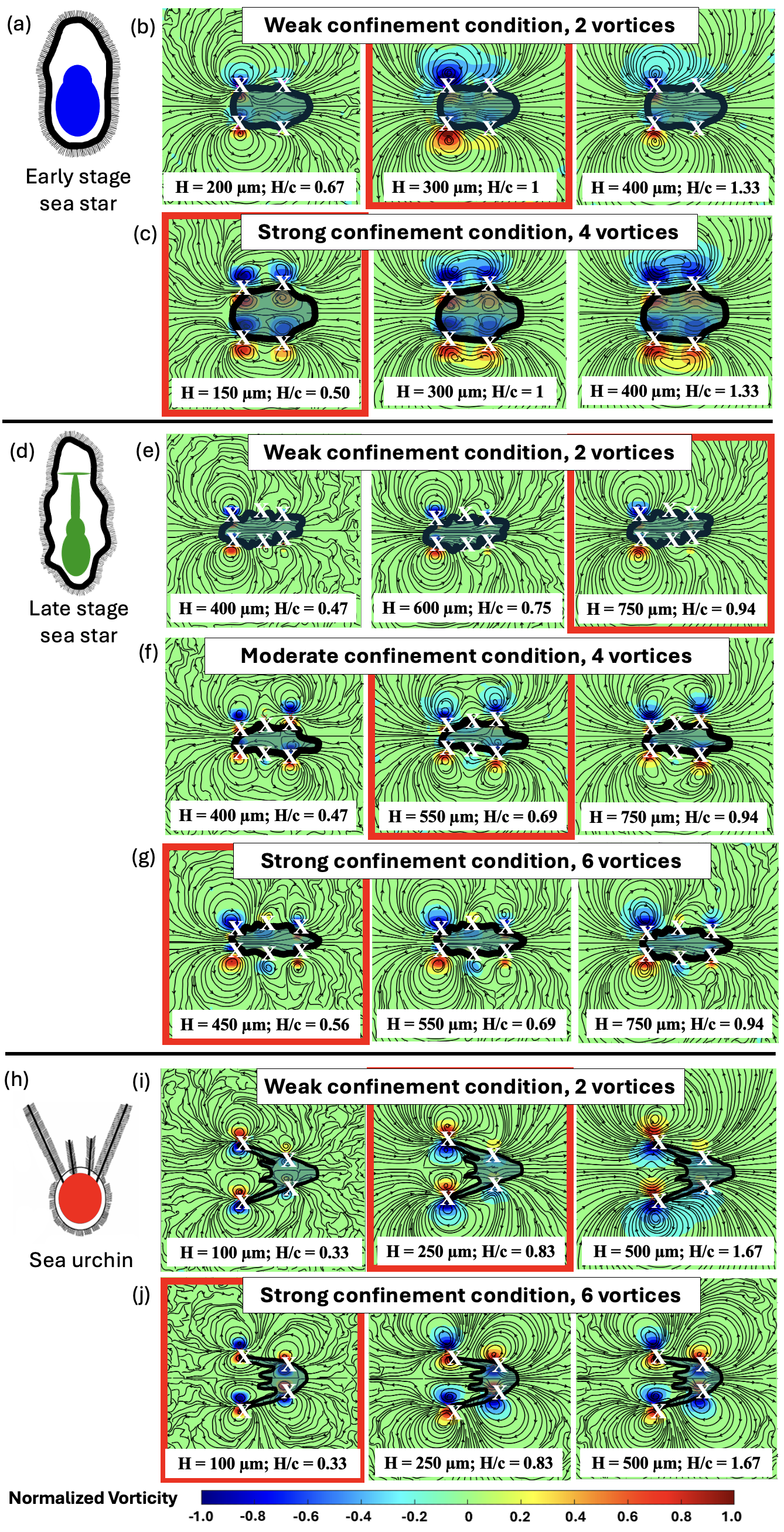}
   \captionsetup{labelformat=empty}
            \caption{\textbf{Extended Data Fig. 4} (Caption next page.)}
		\label{fig:S4}
\end{figure}

%\addtocounter{figure}{-1}
\begin{figure} [t!]
\captionsetup{labelformat=empty}
  \caption{(Previous page.)  \textbf{Extended Data Fig. 4: Theoretical flow-fields around marine larvae under different confinements.} In each panel, the black lines with arrows are streamlines and the heatmap represents vorticity contours.  Theoretical panels closely matching experimental data one-on-one at specific heights are highlighted in red boxes. (a) Early stage sea star larva: Theoretical flow fields under weak and strong confinement. (b) Late stage sea star larva: Theoretical flow fields under weak, moderate, and strong confinement. (b) Sea urchin larva: Theoretical flow fields under weak and strong confinement. The vorticity is normalized for each case (Supplementary Material) and their magnitudes are indicated in the color bar shown below.}
\end{figure}

\begin{figure}
	\centering
		\includegraphics[width=1.1\columnwidth]{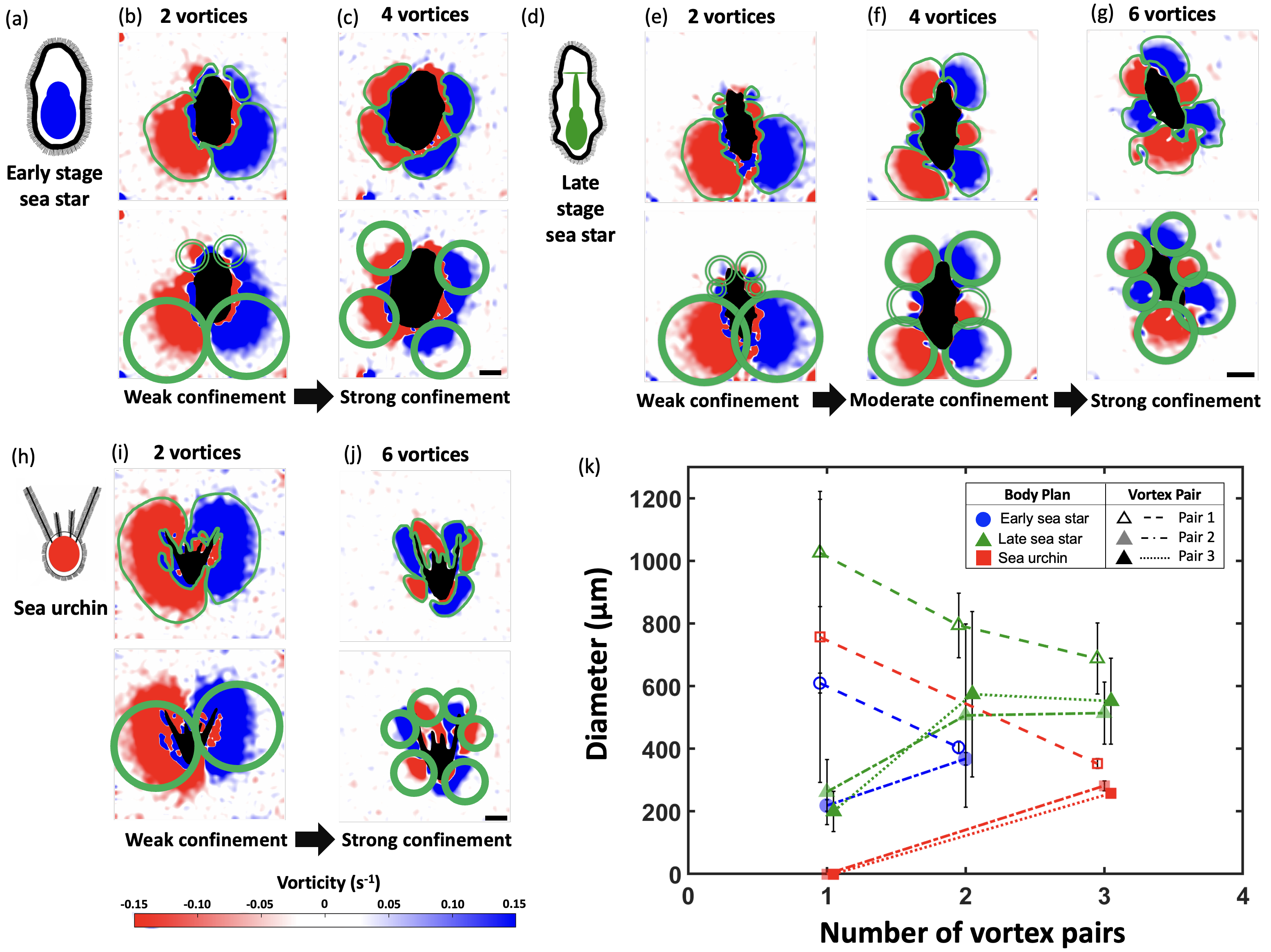}
        \captionsetup{labelformat=empty}
		\caption{\textbf{Extended Data Fig. 5: Quantifying vortex size around marine larvae under different confinements in experiments.} All panels represent the vorticity heatmap (red/blue colorbar below) with 'patches' of vorticity generated by the larvae. Top row: vorticity patches are marked with green boundaries, whose areas are quantified manually. Bottom row: vorticity patches are marked with circles representing the equivalent diameter based on the calculated area from top row. The solid green circles indicate the dominant vortices that can be readily visualized (Figures 1, 2), and the 'dashed/thin' circles represent equivalent vortices that are neither dominant nor seen visually in the experiments. (a) Early stage sea star larva under weak (b) and strong (c) confinements. (d) Late stage sea star larva under weak (e), moderate (f), and strong (g) confinements. (h) Sea urchin larva under weak (i) and strong (j) confinements. (k) The average vortex diameter across a vortex pair and thresholding (Supplementary Table S2), as a function of total number of vortex pairs generated around one case for each body plan (indirectly representing increase in confinement). The dominant Pair 1 vortex size decreases with increase in confinement, whereas Pair 2 and 3 vortices increase in size with increase in confinement. Data points have been jittered about the x-axis for clarity.}
		\label{fig:S5}
\end{figure}

%%%%%%%%%%%%%%%%%%%%%%%%%%%%%%

\end{document}